# Idle Period Propagation in Message-Passing Applications

Ivy Bo Peng, Stefano Markidis, Erwin Laure
Department of Computational Science and Technology
KTH Royal Institute of Technology, Sweden

Gokcen Kestor, Roberto Gioiosa
Computational Science and Mathematics Division
Pacific Northwest National Laboratory, USA

*Abstract*—Idle periods on different processes of Message Passing applications are unavoidable. While the origin of idle periods on a single process is well understood as the effect of system and architectural random delays, yet it is unclear how these idle periods propagate from one process to another. It is important to understand idle period propagation in Message Passing applications as it allows application developers to design communication patterns avoiding idle period propagation and the consequent performance degradation in their applications. To understand idle period propagation, we introduce a methodology to trace idle periods when a process is waiting for data from a remote delayed process in MPI applications. We apply this technique in an MPI application that solves the heat equation to study idle period propagation on three different systems. We confirm that idle periods move between processes in the form of waves and that there are different stages in idle period propagation. Our methodology enables us to identify a self-synchronization phenomenon that occurs on two systems where some processes run slower than the other processes.

## I. INTRODUCTION

One of the most compelling questions in High-Performance Computing (HPC) is how random system noise and small idle periods can propagate among processes of Message Passing (MP) applications and degrade the overall performance of applications. In fact, experimental evidences have shown that random delays occurring over $ns - \mu s$ time scales originate on a single process, propagate among processes and are responsible for the overall performance degradation that can reach $ms$ delays. Despite many studies have been devoted to understanding how noise originates and affects a single process, the mechanisms allowing noise propagate among processes of parallel applications at large-scale are still poorly understood. A full understanding of the noise propagation would allow application developers to avoid noise propagation and consequent performance degradation in designing parallel communication.

The vast majority of parallel scientific applications follow the domain-decomposition strategy. These applications divide the simulation boxes into multiple smaller domains and assign one subdomain workload to a process. For instance, when the heat equation is solved by finite difference technique, the whole simulation box is discretized by imposing a numerical grid in the space and the problem variables are only defined on the grid points. The whole numerical grid is then divided into a set of smaller grids: the variables defined on these smaller grids and the computation of these variables are assigned to different processes. At each computational cycle, each process only updates the variables on its own grid points. However, differential equations include spatial differential operators, such as Laplacian in the heat equation. The calculation of the discretized spatial operator on a grid point requires the values of the variables on the neighbor grid points that define the computational stencil. While the calculation of the discretized spatial differential operator in the inner part of the grid can be performed in parallel by different processes, the calculation of variables on the boundary grid points requires the values of variables on the grids allocated to neighbor processes. In order to obtain these values in MP applications, processes exchange messages for the values of variables at the boundaries.

For this reason, MP applications alternate between computation and communication phases. In the ideal case, if the grid points are equally distributed among domains, each process reaches the communication phase at the same time and exchanges data with other processes. However, it is statistically impossible to have processes progressing at the same computational speed as both hardware and software can have some indeterministic factors impacting a single process, e.g. the temperature of a processor could be a quite dynamic factor during execution. If a process completes the computation phase earlier than the processes it needs to communicate to, it waits for the other processes without performing any calculation. In this case, it is said that the process is in "idle" or there is an "idle period" on the process. While it is straightforward to understand that a single process is affected by an idle period when waiting for other delayed processes, it is less obvious to understand how the same idle period moves to other processes that are not directly connected by communication.

Simulations of MP applications showed that idle periods are generated on a process when it needs to wait for a delayed process. This idle period can propagate among different processes as a wave [1]. An example of such mechanism is shown in Figure 1 that represents an application running on seven processes with data exchange only among neighbor processes. The yellow color represents the computing time while the blue color represents the idle time spent in waiting to complete the communication. In this example, $P0$ is the slowest process and $P1$ waits for $P0$ to complete the computation. Because $P2$ needs to wait for $P1$ to complete the communication with $P0$, $P2$ is also delayed consequently. In the same way, all the other processes $P3 - P6$ wait for the neighbor processes to complete the communication. This communication pattern, involving parallel communication with neighbor processes, results in the propagation of idle periods from $P2$ to $P6$. It has been shown that the propagation speed depends on the average computational time and on network parameters, such as the latency of point-to-point communication [1].

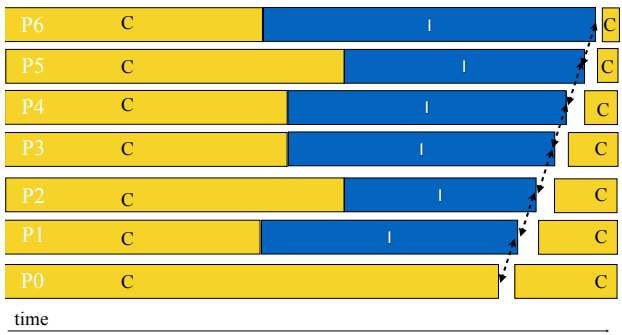

Fig. 1: An idle period that is generated on P1 propagates as an effect of point-to-point communication to the other processes.

Because idle waves originate as an effect of a local (to the process) delay, it is important to understand how a workload imbalance originates. On supercomputers, delays on process are mainly generated in two ways:

1) **Process imbalance**. The time for computing the same workload can be different for different processes. This is statistically unavoidable as processor speed depends on several factors among which the temperature of a processor is one very dynamic factor. In addition, operating system (OS) and architectural noise can suddenly slow down a single process. For instance, the kernel can "interrupt" a process of a parallel applications to execute system-level activities or can "preempt" a process to run other tasks. These delays range from $100\ ns$ (the cost of a cache miss) to $20\ ms$ (the cost of process pre-emption or of a swap-in) [2]. In MP applications, the delay on one process can result in an idle period on the processes waiting for data from this delayed process. This implicit local synchronization of point-to-point communication allows for the propagation of the idle period among communicating processes [1], [3].
2) **Shared resources**. If two or more processes compete for the same resource, some of the processes may be delayed for waiting for the resource to be freed by the other process. For instance, different processes on a compute node might access the memory controller or the Network Interface Controller (NIC) at the same time. If a process tries to access the NIC when it is already in use by another process, a delay will occur on this process until the NIC becomes available.

The goal of this paper is to study how idle periods that are unavoidable on MP applications propagate among different processes. Our main contribution is to introduce a methodology to specifically trace idle periods arising when processes are waiting to complete a communication in MPI applications. This technique can be applied to the study of idle period propagation in any MPI application.

We apply this technique to study idle period propagation carrying out MPI simulations to solve the 1D heat equation on three different systems. By tracing idle period in this application:

1) We confirm experimentally for the first time that idle waves are present in MPI applications on three different systems. Idle waves were predicted by simulation studies of MP applications but never confirmed by experimental studies before.
2) We show different phases of idle period propagation. First, random process imbalance is reinforced on those processes communicating to processes outside the socket or the computing node. Idle waves then propagate idle periods outside the computing nodes.
3) We show for the first time that processes in a MP application can "self-synchronize", exhibiting idle periods occurring with the same frequency but different phases. We show that self-synchronization occurs when a process or a group of processes is considerably slower than other processes.

This paper introduces a methodology to trace the propagation of idle periods in MPI applications and presents its application to studying idle period propagation. The article is organized as follows. Section II describes previous works on idle propagation and system noise. Section III introduces the methodology to trace idle period propagation and the experimental set-up. We show how to use this methodology to study idle period propagation in Section IV, focusing on idle period propagation and self-synchronization of different processes. Finally, we summarize the results and conclude the paper in Section V.

## II. RELATED WORK

Previous works on simulations of idle period propagation showed that idle period propagates among processes in certain point-to-point communication patterns such as neighbor communication [1] and collective operations [3]. LogGOPSim simulations [4] predicted that idle periods propagate among MPI processes as waves with a propagation speed depending on the average computational time and network parameters [1]. Montecarlo simulations showed that certain classes of collective operations have the property of hiding process imbalance and minimizing idle period propagation [3]. Different from these studies, we experimentally trace the idle period in an MPI application to study the idle period propagation on three different systems.

Idle period and its impact on parallel applications have been extensively studied via various techniques, including noise injection [5] and simulation [6]. In the HPC community, Petrini et al. [7] explained how operating system (OS) noise and other system activities could drastically impact the performance of a large cluster. In the same work, they observed that the impact of the system noise when scaling on 8K processors was largely due to *noise resonance* and that leaving one processor idle to take care of the system activities brought a $1.87\times$ performance improvement. In a following work [8], the authors identified timer interrupts, and the activities started by the paired interrupt handler, as the main source of system noise. Others [9], [10] found the same result using different methodologies. The other major cause of system noise is the kernel scheduler as the kernel can swap HPC processes out in order to run other processes, including kernel daemons. This problem has also been extensively studied [7], [8], [9], [10], [11], [12], and several solutions are available [13], [12].

In a different work based on simulation results, [6] pointed out that both collective and point-to-point operations can impact the sensitivity of parallel applications to system noise and that the blocking communication mode has an amplification effect of system noise. Also based on simulation results, [1] presented an equation description of the propagation speed of idle waves in scientific applications on large number of processes.

Studies categorize OS noises into high-frequency, short-duration noise (e.g., timer interrupts) and low-frequency, long-duration noise (e.g., kernel threads) [5]. Impact on HPC applications is higher when the OS noise resonates with the application, i.e., high-frequency, fine-grained noise affects more fine-grained applications, and low-frequency, coarse-grained noise affects more coarse-grained applications [7], [5]. The impact of the operating system on classical MPI operations, such as collective, is examined in Beckman et al. [2]. Morari et al. [14] analyze the entire Linux kernel by instrumenting all kernel entry points and fundamental activities, thus providing a complete OS noise analysis. They were able to measure events like page faults and soft interrupt requests that significantly contribute to OS noise that were not measured in previous work.

This extensive body of work on system noise lead to the development of several micro and lightweight kernels (LWKs). Examples of micro-kernels include *L4* [15], *Exokernel* [16], [17], [18], and *K42* [19]. Sandia National Laboratories has a history of developing lightweight kernels dating back to the early 90s. Riesen et al. [20] described the evolution, design and performance of Sandia's LWKs, highlighting the benefits of their implementations over standard OSes like Linux or OSF/1. The authors identified predictable performance and scalability as major goals for HPC systems. IBM *Compute Node Kernel* for Blue Gene supercomputers [21], [22] is an example of a LWK targeted for HPC systems. CNK is a standard open-source, vendor-supported, OS that provides maximum performance and scales to hundreds of thousands of nodes. There are also several full-weight HPC kernels, including ZeptoOS [23], [24], [25], IBM AIX [11], and various other Linux variants [12].

## III. METHODOLOGY AND EXPERIMENTAL SET-UP

The main contribution of this paper is to introduce a methodology to trace idle period propagation in MPI applications and apply it to trace a real application on three different systems. This section presents this methodology, the MPI application and the experimental set-up.

### A. Tracing Idle Periods in MPI Communication

To study the propagation of idle periods, we need to first measure the time a single process is idling while waiting to receive data from a remote process and then reconstruct the propagation of these idles periods among all processes. In our work, the time spent in waiting for incoming data (idle time) is measured by the time spent in waiting for a receive operation to complete. While it is straightforward to measure the idle time in non-blocking MPI operations, where communication operations starts with `MPI_Irecv` and complete exiting `MPI_Wait`, it is not easy to directly measure the idle time spent in blocking MPI communication at application level. For this reason, we have developed a MPI wrapper library acting as an interposition library that automatically splits calls to `MPI_Recv` into the semantically equivalent `MPI_Irecv` + `MPI_Wait` and measure the number clock cycles elapsed in waiting for a receive operation to complete. No other operation is performed between these two operations except for starting a timer before the `MPI_Wait` function. We measure the idle time as the number of clock cycles elapsed between the entry and exit of the `MPI_Wait` function. The number of cycles spent waiting is measured using the Time Stamp Counter (TSC) and the instruction `RDTSC`.

This technique excludes the time spent in posting a receive request from MPI blocking receive operation, thus provides a more accurate profile of the idle time on each process. To quantify the overhead of `MPI_Irecv` + `MPI_Wait` over the original `MPI_Recv`, we compared the wall time of running 100 computational cycles on 256 processes using these two approaches and find less than 0.01% overhead. We collect the trace of each process running the application and reconstruct the idle period propagation from all the traces.

We note that it would be possible to use internal MPI performance counters to identify the time spent in polling the message queue. The access to these counters is now possible via the new `MPI_T` tool interface that has been introduced in the MPI-3 standard [26]. However, at the time of writing this paper, MPI implementations do not provide such access to the performance variables yet. We plan to investigate such interface for tracing idle periods in the future and compare the results with our approach.

### B. Heat MPI Application

We selected an MPI application for the solution of the heat equation in one dimensional domain decomposition using finite difference discretization in space and explicit discretization in time. The structure of the application is fairly simple: each process sends and receives data from its two neighbors at each computational cycle. The application is written in *C* and it uses MPI blocking point-to-point communication. The code can be downloaded from [27].

The same amount of workload is assigned to each MPI process in this benchmark, so that imbalance is attributed to the noise other than the workload imbalance. We keep a constant number of grid points $N_G = 10^6$ per process in the experiments. All the tests were carried out using the same set of default parameters on all three systems. We also varied the total number of computational cycles from 100 to 1,000 and 10,000 to study idle period propagation at different time scales. Each process exchanges data with its left and right neighbors at each computational cycle. For this reason, there are two measured idle periods per computational cycles with potentially different characteristic idle periods.

In this paper, we show results that have been obtained running 256 MPI processes. Additional tests have been run on smaller and larger number of processes confirming the results obtained on 256 processes. We repeated the tests five times to verify the consistency of the results and found excellent agreements among all the results.

*C. Test Environment*

We carried out experiments on three systems with different hardware (interconnection network and processors), software ($C$ compilers and MPI implementations) and scale (number of cores per compute node and total number of nodes).

1) The Pal cluster at the Pacific Northwest National Laboratory is a x86 AMD commodity cluster that uses 2.1 GHz AMD Interlagos processors and Mellanox ConnectX-2 InfiniBand network. The cluster has a total of 128 nodes and each node has 32 cores divided between two sockets, with 16 cores on each. Applications are compiled with GNU C compiler version 4.6.2 and the OpenMPI library version 1.8.4.
2) The Seapearl cluster at the Pacific Northwest National Laboratory consists of 32 Ivy bridge nodes connected with 4xQDR Infiniband. Each node has two Intel Xeon E5-2680v2 10-core (2.8 GHz) processors on two sockets. Applications are compiled with GNU C compiler version 4.4.7 and the OpenMPI library version 1.8.4. Both clusters are running on GNU/Linux operating system.
3) Beskow at KTH is a Cray XC40 system with Intel Xeon E5-2698v3 16-core (2.3 GHz) processors and Cray Aries interconnect network with Dragonfly topology. It has a total of 1,676 compute nodes: each node has 32 cores divided between two sockets with 16 cores on each. The operating system is Cray Linux, and the applications are compiled with the Cray C compiler version 5.2.40 with optimization flag -O3 and the Cray MPICH2 library version 7.0.4.

The Pal and Beskow systems are respectively the oldest (slowest) and newest (fastest) among the three systems under study. Table 1 summarizes the hardware and software for the three different experimental set-up.

TABLE I: Specification of Test Systems

| System | Pal | Seapearl | Beskow |
|---|---|---|---|
| Processor | AMD Interlagos | Intel IvyBridge | Intel Haswell |
| Interconnect | Mellanox | 4xQDR | Aries |
| Compiler | Cray C v5.2.40 | GNU C v4.6.2 | GNU C v4.4.7 |
| MPI | OpenMPIv1.8.4 | OpenMPIv1.8.4 | MPICH2v7.0.4 |

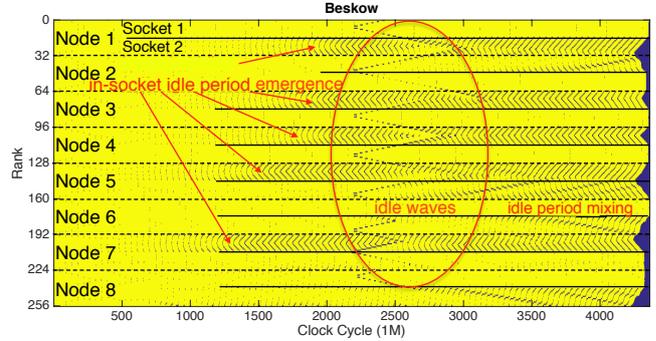

Fig. 2: The three phases of idle period propagation. Idle periods appear first on some processes within one socket. Idle waves then propagate the idle periods within one socket. Finally, idle waves mix idle periods outside the socket.

## IV. RESULTS

We apply the described methodology to trace and reconstruct idle period propagation in an MPI application solving the heat equation on three different systems. We analyze the traces with idle periods and identify the characteristics of idle period propagation among MPI processes. Furthermore, we found evidence of process self-synchronization in two systems under study.

*A. Idle Period Propagation*

To understand how the idle periods propagate, we study a short execution of the heat application for 100 computational cycles on 256 MPI processes on the Beskow supercomputer. Figure 2 shows the traces in a contour-plot with time (in clock cycles) on the $x$ axis and with the MPI rank on the $y$ axis. The blue color indicates the idle time that a process spends in waiting for incoming data (idle period). Idle periods that last less than 1M clock cycles are not represented to ease the visualization. The yellow color indicates the time a process is working on some other task, e.g. computation. This plot allows us to understand how idle periods (blue color) propagate both in time ($x$ direction) and among ranks ($y$ direction).

The traces of the idle period propagation on the Beskow supercomputer reveal clear dependence on the underlying hardware topology in the initial stage. It is noted that the 256 MPI processes are distributed among eight Beskow compute nodes. Each compute node has two sockets with 16 processes. This layout of 16 processes within one single socket can be clearly identified from Figure 2 in the form of a wave propagating in one socket in one compute node. In Figure 2, we superimpose dashed lines to separate the processes belonging to different nodes. We also superimpose solid lines to separate processes belonging to different sockets within a node.

Three different phases in the idle period propagation are visible when inspecting Figure 2:

1) **In-socket idle period emergence**. Idle periods appear only on a group of processes relative to one socket on a compute node. This is due to the positive feedback to random noises: the initial random noise on some processes is reinforced at each computational cycle. For instance, random delays on processes competing for shared resources are reinforced at the next computational time as the delayed processes are likely to access the shared resources late again. In the application under study, idle period is only reinforced on two processes within one socket. This is clear from Figure 2: the growth of idle time is evident within the socket 2 in node 1. In particular, the delays on two processes (one communicating out of the node and one communicating out of the socket) are amplified at each computational cycle. However, it is not clear at a first analysis how the idle periods that are only present on the processes within one socket propagate to processes on other sockets and other compute nodes.

2) **Idle waves**. Idle periods within one socket propagate to other processes that were not affected by long idle periods in the first stage. This propagation is in the form of two idle waves moving in opposite directions. Such idle waves can be generated by a sudden large delay on one process. For instance, the cause of sudden large delays could be some large amplitude OS noise or other system operations. These idle waves are visible as oblique blue lines with opposite slopes in the dashed red ellipse of Figure 2.

3) **Idle period mixing**. Idle waves propagate the idle periods to processes running on different sockets. In this phase, if the processes on different sockets have no pre-existing long idle periods, the idle waves bring idle periods to these processes. Otherwise, if there are already idle periods existing on the processes, the idle waves interact with those pre-existing idle periods. This last phase is clear when inspecting the contour plot in the nodes 5 and 6: the interaction between idle waves and existing idle periods appear as blue stripes.

We observe idle period propagation on all the three test systems. A direct comparison of the idle period dynamics on the three systems is presented in Figure 3. This figure visualizes the traces of the benchmark applications running 1,000 computational cycles. We chose a longer execution because we are interested to study idle period dynamics during the lifetime of an application. The three inserts in Figure 3 show the initial period of the simulation. When inspecting the inserts, the emergence of in-socket idle period is evident in all the three systems. To complete the simulation on the three systems took approximately between 45 G (Beskow), 60 G (Seapearl) and 95 G (Pal) clock cycles, roughly corresponding to 45, 60 and 95 seconds of execution times.

We first study idle period propagation on the Pal supercom-puter. The top panel of Figure 3 shows the propagation of idle waves at several points of the simulation. The largest noise amplification occurs at 30 G clock cycles when a series of large amplitude idle waves propagate to the whole system. Such large amplitude occurred only once during this simulation of 1,000 computational cycles. However, we observed multiple large amplitude idle waves during a simulation with 10,000 computational cycles (results are not shown in this paper). Besides the large amplitude idles waves, smaller scale idle waves are also clear in the simulation, i.e. at 15 G clock cycles. The propagation speed of the idle waves can be calculated by evaluating the slope of the black lines at 15 G clock cycles. Their speed is approximately 128 MPI ranks in 2 G clock cycles, which is roughly 64 ranks per second.

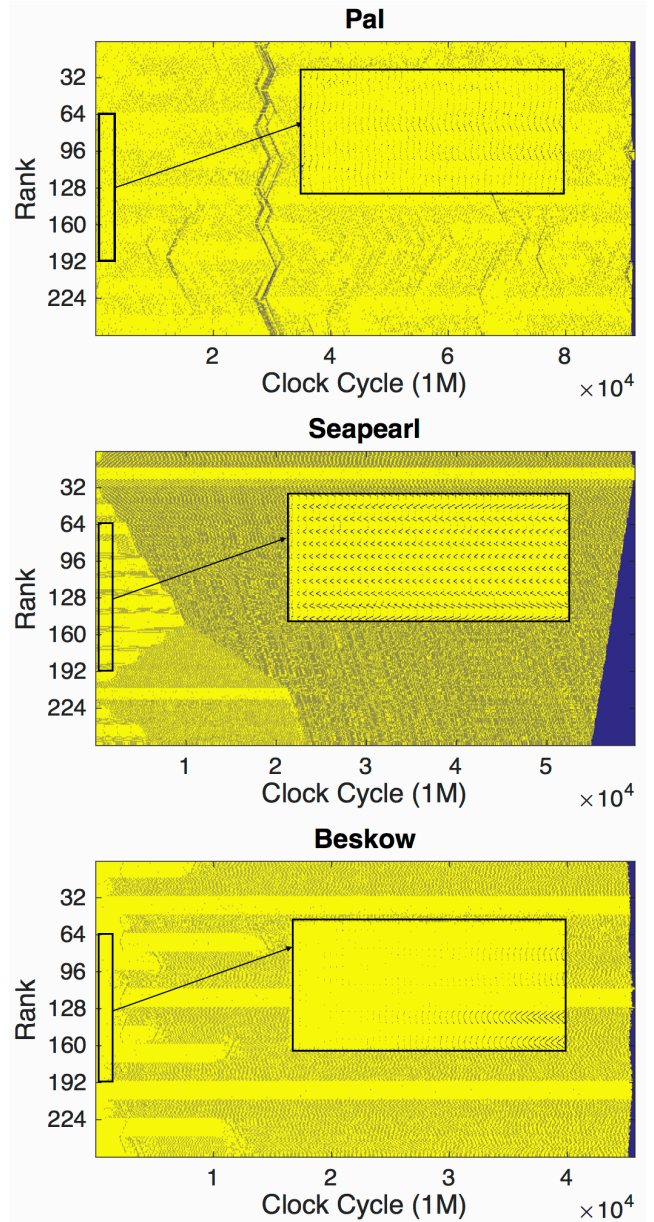

Fig. 3: Contour plot of the idle time (blue color) on 256 processes running 1,000 computational cycles of the heat application on the three systems. The $x$ axis indicates the time in clock cycles and the $y$ axis indicates the MPI rank. The inserts show the initial part of the simulation.

We then study idle period propagation on the Seapearl supercomputer. The middle panel of Figure 3 shows the contour plot of idle periods on the Seapearl system. We notice that the parallel application exhibits large coherent idle periods that affect all the ranks during the simulation. These large coherent structures of idle periods occur on almost all the ranks after 25 G clock cycles. The only exception is a set of 10 processes (a yellow band at the top of the contour plot). These are the slowest group of processes and do not present any large idle period. It is noted that each socket consists of 10 processes on the Seapearl supercomputer and these slowest processes belong to the same socket.

Finally, we analyze idle periods on Beskow. The contour plot of idle periods on Beskow also shows the emergence of in-socket idle time and its propagation by idle times (Figures 2 and 3). Note that the bottom panel of Figure 3 visualizes the traces of $1,000$ computational cycles while Figure 2 shows the simulation results only up to 100 computational cycles. The two plots show a similar dynamics for the idle periods. The in-socket idle periods start independently on 16 consecutive ranks corresponding to one socket. At approximately 10 G clock cycles, idle periods start merging and clustering into four large groups of idle periods till the end of execution of the application. Additional simulations with ten times more computational cycles show that eventually these regions with large idle period merge into one and the large structure of coherent idle periods spreads to the whole system and persists till the end of simulation.

*B. Self-Synchronization in MPI Applications*

An analysis of idle periods on different processes of Seapearl and Beskow systems points out that long idle periods appear as coherent structures on both systems after a certain number of clock cycles. For instance, Figure 4 and the middle panel of Figure 3 provide examples of the appearance of such coherent structures. The initial stage of random idle period rapidly transitions to structured long idle periods. These structures occur on all the processes except for the group of slowest processes (the yellow band in Figure 4). We find that this phenomenon occurrs on different simulations on the same machines. For instance, Figure 4 and the middle panel of Figure 3 are based on two different runs on the Seapearl system. Despite the details of the temporal evolution of the two runs are different, i.e the group of slowest processes is different, the overall macroscopic evolution of the system is very similar. In fact, coherent structures of long idle periods are present on all the processes except for the slowest ones after a certain period. An insert in Figure 4 shows the details during the period from 0 to 15 G clock cycles.

Coherent structures of idle periods occur on a group of ranks corresponding to the processes relative to one socket

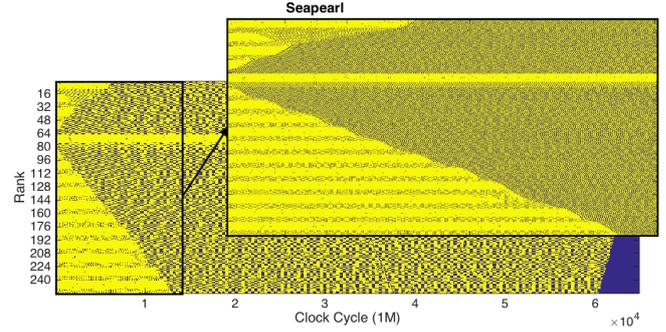

Fig. 4: Contour plot of the idle time (blue color) with clock cycles on the $x$ axis and the MPI rank on $y$ axis. The heat application runs $1,000$ computational cycles on the Seapearl system. The insert shows a zoom-in of the transition phase to self-synchronization of MPI processes.

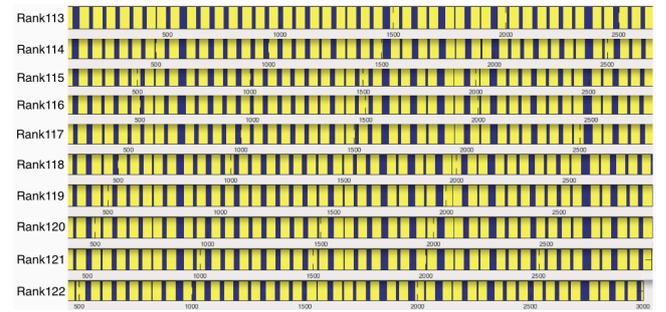

Fig. 5: Computing (in yellow) and idle (in blue) periods for MPI ranks $113 - 122$ on Seapearl after self-synchronization occurs. The time in cycles is shifted for each process to show that different processes have the same idle periods but at different phases.

(10 processes). It is clear in the contour plot that these groups of processes develop prolonged idle periods indicated as blue bands. These coherent structures of idle periods propagate to the whole system. It is clear from the insert in Figure 4 that this phenomenon originates from rank 64 to 73 and it propagates to other ranks: it either brings large organized idle time to those ranks previously free of local idle period or it merges with the pre-existing idle period. We also notice that the propagation speed of this phenomenon is almost constant (the blue line delimiting the coherent structure in the time-rank plot). This propagation velocity is approximately 176 ranks per 15 G clock cycles, which is roughly 12 ranks per second.

However, the nature of the idle period coherent structures is unclear by simply inspecting the contour plot of Figure 4. To understand this, we plot the idle period structures for different contiguous processes (ranks $113 - 122$) shifting the time axis for each process by a small number of clock cycles in Figure 5. By introducing this time shift, it is clear that each process shows exactly the same idle period pattern but at different phases. After a certain number of computational cycles, different MPI processes of the heat application self-synchronize, exhibiting the same idle period pattern but at different phases (or time shift).

We notice that self-synchronization occurs on the Seapearl and Beskow systems but it seems to be absent in the simulations on the Pal supercomputer. To understand the reason for process self-synchronization occuring only on the Seapearl and Beskow systems, we study the characteristics of the idle periods and plot the minimum, maximum and average idle periods on the three systems in Figure 6. Despite the three test environments have different processors, network and software, they have very similar characteristics of idle periods. In fact, on all the three systems, the minimal time spent in waiting for data is between $100$ and $1,000$ clock cycles (100 ns - $1\mu$s), the average is $10^6 - 10^7$ clock cycles (1 ms - 10 ms) and the maximum is $10^7 - 10^8$ clock cycles (10 ms -100 ms). However, we notice that the heat application on the Pal supercomputer does not have a group of processes running considerably slower than other: the average idle period is approximately the same for all ranks. A different situation occurs on the Seapearl and Beskow supercomputer, where a group of processes show a considerably smaller average idle period (corresponding to longer computational phase and therefore to slower processes). The ranks $64 - 73$ and $0 - 15$ are slowest by one order of magnitude respectively on the Seapearl and Beskow supercomputers. For this reason, it is likely that self-synchronization occurs only on systems where a group of considerably slower processes are present.

## V. DISCUSSION AND CONCLUSIONS

It is an experimental fact that idle periods occur on different processes of MP applications and propagate to other processes. However, the mechanism enabling idle period propagation among processes and its overall effects on MP applications are still unclear.

In this work, we introduced a methodology to trace idle periods in MPI applications and we applied it to study idle period propagation in the heat application that uses MPI blocking point-to-point communication. By tracing idle periods, we analyzed idle propagation on three different systems with different hardware and software. We found that the processes self-synchronize on simulations/systems that present a group of processes considerably slower than other ones. Our methodology allows application developers to understand how noise propagates in MPI applications and guides them to design parallel communication pattern that can avoid noise propagation and its consequent performance degradation.

The methodology that we presented can be applied to any MPI application. However, this paper shows only the study of idle period propagation in blocking MPI communication. Non-blocking MPI communication allows for overlapping of communication and computation by providing a two-stage communication: communication is initiated first, i.e. `MPI_Irecv` and the completion of the communication is only ensured at a later stage i.e. `MPI_Wait`. For this reason idle time can be hidden by as much as the time of computation between the two stages of communication in the non-blocking communication mode. In real-world applications, it is not always feasible to achieve a complete overlapping of computation and communication.

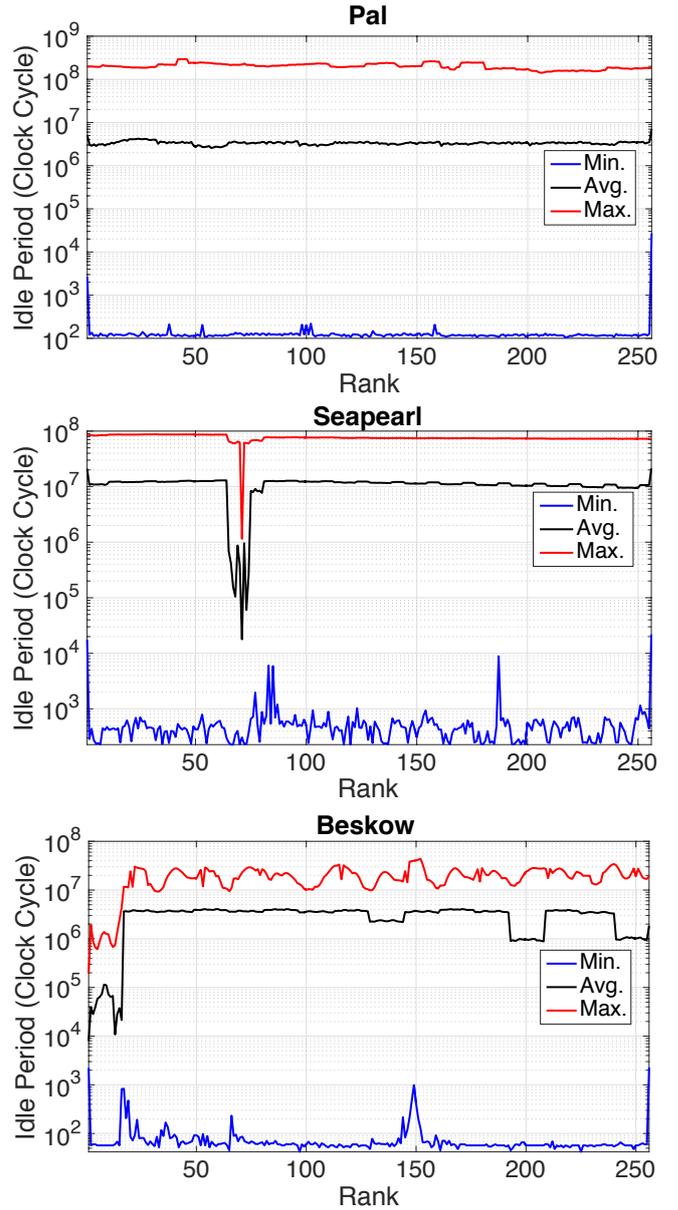

Fig. 6: The minimum, average and maximum idle period on each rank on the three systems are presented.

In such case, idle period can still be generated. Overall, non-blocking communication is less affected by idle period propagation and idle period can be partially or fully absorbed by computation depending on the application.

In message-passing systems, collective operations are implemented atop several point-to-point communications. For this reason, blocking collective operations are also affected by noise amplification [6]. Future work will focus on investigating idle period propagation in collective operations following our methodology: blocking collective operations are split into non-blocking collectives and one synchronization point (`MPI_Wait`).

We showed that self-synchronization took a certain number of cycles to fully develop from the random delays to coherent idle period structures. For this reason, it is important to carry out long enough experiments to observe self-synchronization of MPI processes. We note that the Pal system is an older generation system while the Seapearl and Beskow are newer supercomputers. One important point is that the older generation systems (slower network, slower CPU processor) show an average idle period that is of the same order of the newer systems. This indicates that idle period propagation is still an import problem to solve in applications running on current and future supercomputers.

ACKNOWLEDGMENT

This work was funded by the European Commission through the EPiGRAM (grant agreement no. 610598) and SAGE projects (grant agreement no. 671500). This work was supported by the DOE Office of Science, Advanced Scientific Computing Research, under the Argo project (award number 66150) and the CENATE project (award number 64386). This research used computing resources provided by the Swedish National Infrastructure for Computing (SNIC) at PDC.